\newif\ifapsmode
\apsmodetrue 

\ifapsmode
\documentclass[prl,twocolumn,twoside,superscriptaddress,nofootinbib]{revtex4}
\else
\documentclass[12pt]{nature}
\usepackage[margin=1in]{geometry}
\fi

\usepackage{graphicx} 
\usepackage[pdfauthor={Erik Lucero}, pdftitle={Computing prime factors with a Josephson phase qubit quantum processor}]{hyperref} 
\usepackage{multirow} 
\usepackage{dcolumn} 

\graphicspath{{/}}

\newcommand{\ket}[1]{\ensuremath{\left|#1\right\rangle}}
\newcommand{\bra}[1]{\ensuremath{\left\langle#1\right|}}

\newcommand{\GHZ}{\ket{\mathrm{GHZ}}}

\newcommand{\W}{\ket{\mathrm{W}}}

\renewcommand{\Im}{\ensuremath{\mathrm{Im}}}

\newcommand{\UCSB}{Department of Physics, University of California, Santa Barbara, CA 93106, USA}

\begin{document}

\title{Computing prime factors with a Josephson phase qubit quantum processor}

\ifapsmode
\author{Erik Lucero}\affiliation{\UCSB}
\author{R. Barends}\affiliation{\UCSB}
\author{Y. Chen}\affiliation{\UCSB}
\author{J. Kelly}\affiliation{\UCSB}
\author{M. Mariantoni}\affiliation{\UCSB}
\author{A. Megrant}\affiliation{\UCSB}
\author{P. O'Malley}\affiliation{\UCSB}
\author{D. Sank}\affiliation{\UCSB}
\author{A. Vainsencher}\affiliation{\UCSB}
\author{J. Wenner}\affiliation{\UCSB}
\author{T. White}\affiliation{\UCSB}
\author{Y. Yin}\affiliation{\UCSB}
\author{A. N. Cleland}\affiliation{\UCSB}
\author{John M. Martinis}\affiliation{\UCSB}
\fi

\begin{abstract}
A quantum processor (QuP) can be used to exploit quantum mechanics to find the prime factors of composite numbers\cite{Peter1994}. Compiled versions of Shor's algorithm have been demonstrated on ensemble quantum systems\cite{Vandersypen2001} and photonic systems\cite{Lanyon2007,Lu2007,Politi2009}, however this has yet to be shown using solid state quantum bits (qubits). Two advantages of superconducting qubit architectures are the use of conventional microfabrication techniques, which allow straightforward scaling to large numbers of qubits, and a toolkit of circuit elements that can be used to engineer a variety of qubit types and interactions\cite{Clarke2008, Mariantoni2011}. Using a number of recent qubit control and hardware advances \cite{Hofheinz2008, Ansmann2009, Neeley2010, DiCarlo2010, Altomare2010, Yamamoto2010, Mariantoni2011}, here we demonstrate a nine-quantum-element solid-state QuP and show three experiments to highlight its capabilities. We begin by characterizing the device with spectroscopy. Next, we produces coherent interactions between five qubits and verify bi- and tripartite entanglement via quantum state tomography (QST) \cite{Hofheinz2008, Fink2009, Altomare2010,Steffen2006a}. In the final experiment, we run a three-qubit compiled version of Shor's algorithm to factor the number 15, and successfully find the prime factors $48\,\%$ of the time. Improvements in the superconducting qubit coherence times and more complex circuits should provide the resources necessary to factor larger composite numbers and run more intricate quantum algorithms.
\end{abstract}
\ifapsmode
\maketitle
\fi


In this experiment, we scaled-up from an architecture initially implemented with two qubits and three resonators \cite{Mariantoni2011} to a nine-element quantum processor (QuP) capable of realizing rapid entanglement and a compiled version of Shor's algorithm. The device is composed of four phase qubits and five superconducting coplanar waveguide (CPW) resonators, where the resonators are used as qubits by accessing only the two lowest levels. Four of the five CPWs can be used as quantum memory elements as in Ref. \cite{Mariantoni2011} and the fifth can be used to mediate entangling operations. 

The QuP can create entanglement and execute quantum circuits\cite{Nielsen2000, Barenco1995} with high-fidelity single-qubit gates ($X$, $Y$, $Z$, and $H$), \cite{Lucero2008, Lucero2010}combined with  swaps and controlled-phase ($C_{\mathrm{\phi}}$) gates\cite{DiCarlo2009, Yamamoto2010, Mariantoni2011}, where one qubit interacts with a resonator at a time. The QuP can also utilize ``fast-entangling logic" by bringing all participating qubits on resonance with the resonator at the same time to generate simultaneous entanglement\cite{Tessier2003}.  At present, this combination of entangling capabilities has not been demonstrated on a single device. Previous examples have shown: spectroscopic evidence of the increased coupling for up to three qubits coupled to a resonator\cite{Fink2009}, as well as coherent interactions between two and three qubits with a resonator\cite{Altomare2010}, although these lacked tomographic evidence of entanglement. 

Here we show coherent interactions for up to four qubits with a resonator and verify genuine bi- and tripartite entanglement including Bell \cite{Ansmann2009} and $\W$-states \cite{Neeley2010} with quantum state tomography (QST). This QuP has the further advantage of creating entanglement at a rate more than twice that of previous demonstrations\cite{Altomare2010, Neeley2010}. 



In addition to these characterizations, we chose to implement a compiled version of Shor's algorithm\cite{Beckman1996, Buscemi2011}, in part for its historical relevance\cite{Nielsen2000} and in part because this algorithm involves the challenge of combining both single- and coupled-qubit gates in a meaningful sequence. We constructed the full factoring sequence by first performing automatic calibration of the individual gates and then combined them, without additional tuning, so as to factor the composite number $\mathrm{N}=15$ with co-prime $\mathrm{a} =4$, (where $1<\mathrm{a}<N$ and the greatest common divisor between $\mathrm{a}$ and $\mathrm{N}$ is $1$). We also checked for entanglement at three points in the algorithm using QST.

Figure \ref{sample}a shows a micrograph of the QuP, made on a sapphire substrate using Al/AlO$_x$/Al Josephson junctions. Figure \ref{sample}b shows a complete schematic of the device. Each qubit $\mathrm{Q_{i}}$ is individually controlled using a bias coil that carries dc, rf- and GHz-pulses to adjust the qubit frequency and to pulse microwaves for manipulating and measuring the qubit state. Each qubit's frequency  can be adjusted over an operating range of $\sim 2\,\rm{GHz}$, allowing us to couple each qubit to the other quantum elements on the chip. Each $\mathrm{Q_i}$ is connected to a memory resonator $\mathrm{M_i}$, as well as the bus B, via interdigitated capacitors. Although the coupling capacitors are fixed, Fig.\,\ref{sample}c illustrates how the effective interaction can be controlled by tuning the qubits into or near resonance with the coupling bus (coupling ``on'') or detuning $\mathrm{Q_i}$ to $f_B\pm 500\,\mathrm{MHz}$ (coupling ``off'')\cite{Hofheinz2009}. 

\ifapsmode
\begin{figure}
\else
\begin{figure}
\fi
\includegraphics[scale=1.0, trim=0in 0in 0in 0in, clip=false]{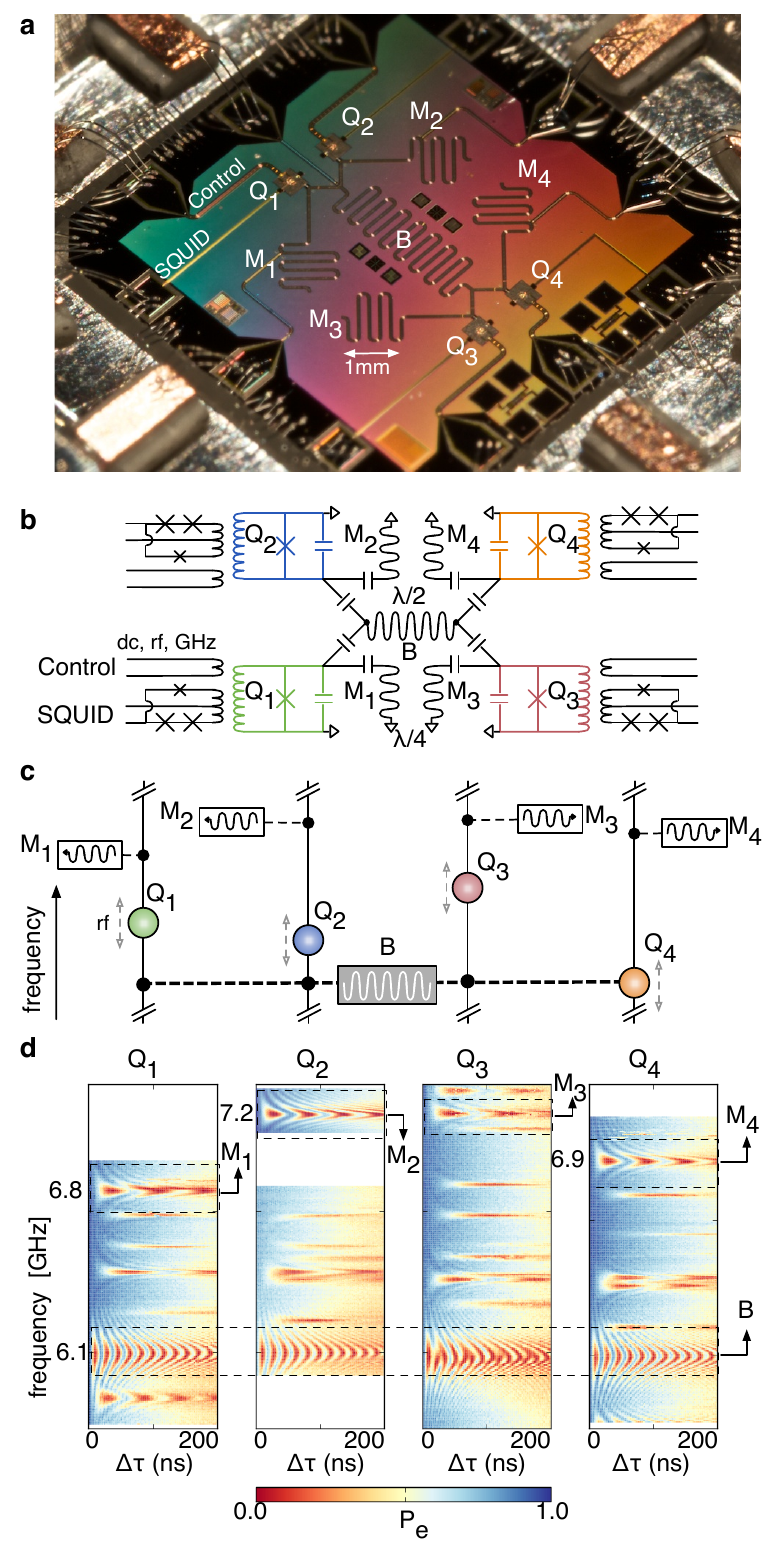}
\caption{\label{sample} Architecture and operation of the quantum processor (QuP). {\bf a}, Photomicrograph of the sample, fabricated with aluminum (colored) on sapphire substrate (dark). {\bf b}, Schematic of the QuP.  Each phase qubit $\mathrm{Q_i}$ is capacitively coupled to the central half-wavelength bus resonator B and a quarter-wavelength memory resonator $\mathrm{M_i}$. The control lines carry GHz microwave pulses to produce single qubit operations. Each $\mathrm{Q_i}$ is coupled to a superconducting quantum interference device (SQUID) for single-shot readout. {\bf c},  Illustration of QuP operation. By applying pulses on each control line, each qubit frequency is tuned in and out of resonance with B (M) to perform entangling (memory) operations. {\bf d}, Swap spectroscopy\cite{Mariantoni2011} for all four qubits: Qubit excited state $\ket{\mathrm{e}}$ probability $P_{\mathrm{e}}$ (color scale) versus frequency (vertical axis) and interaction time $\Delta\tau$. The centers of the chevron patterns gives the frequencies of the resonators B, $\mathrm{M_1}-\mathrm{M_4}$, $f=6.1,6.8,7.2,7.1,6.9\,\mathrm{GHz}$ respectively. The oscillation periods give the coupling strengths between $\mathrm{Q_{i}}$ and B ($\mathrm{M_{i}}$), which are all $\cong 55\,\mathrm{MHz}$ ($\cong 20\,\mathrm{MHz}$).}
\ifapsmode
\end{figure}
\else
\end{figure}
\fi

The QuP is mounted in a superconducting aluminum sample holder and cooled in a dilution refrigerator to $\sim 25\,\mathrm{mK}$. Qubit operation and calibration are similar to previous works\cite{Hofheinz2008, Ansmann2009, Neeley2010, Yamamoto2010, Mariantoni2011},with the addition of an automated calibration process\cite{Lucero2012a}. As shown in Fig.\ref{sample}d, we used swap spectroscopy\cite{Mariantoni2011} to calibrate all nine of the engineered quantum elements on the QuP, the four phase qubits ($\mathrm{Q_1}-\mathrm{Q_4}$), the four quarter-wave CPW quantum memory resonators ($\mathrm{M_{1}}-\mathrm{M_{4}}$), and one half-wave CPW bus resonator (B). The coupling strengths between $\mathrm{Q_i}$ and B ($\mathrm{M_i}$) were measured to be within $5\,\%$ ($10\,\%$) of the design values.



\ifapsmode
\begin{figure*}
\else
\begin{figure}
\fi
\includegraphics[scale=1.0, trim=0in 0in 0in 0in, clip=false]{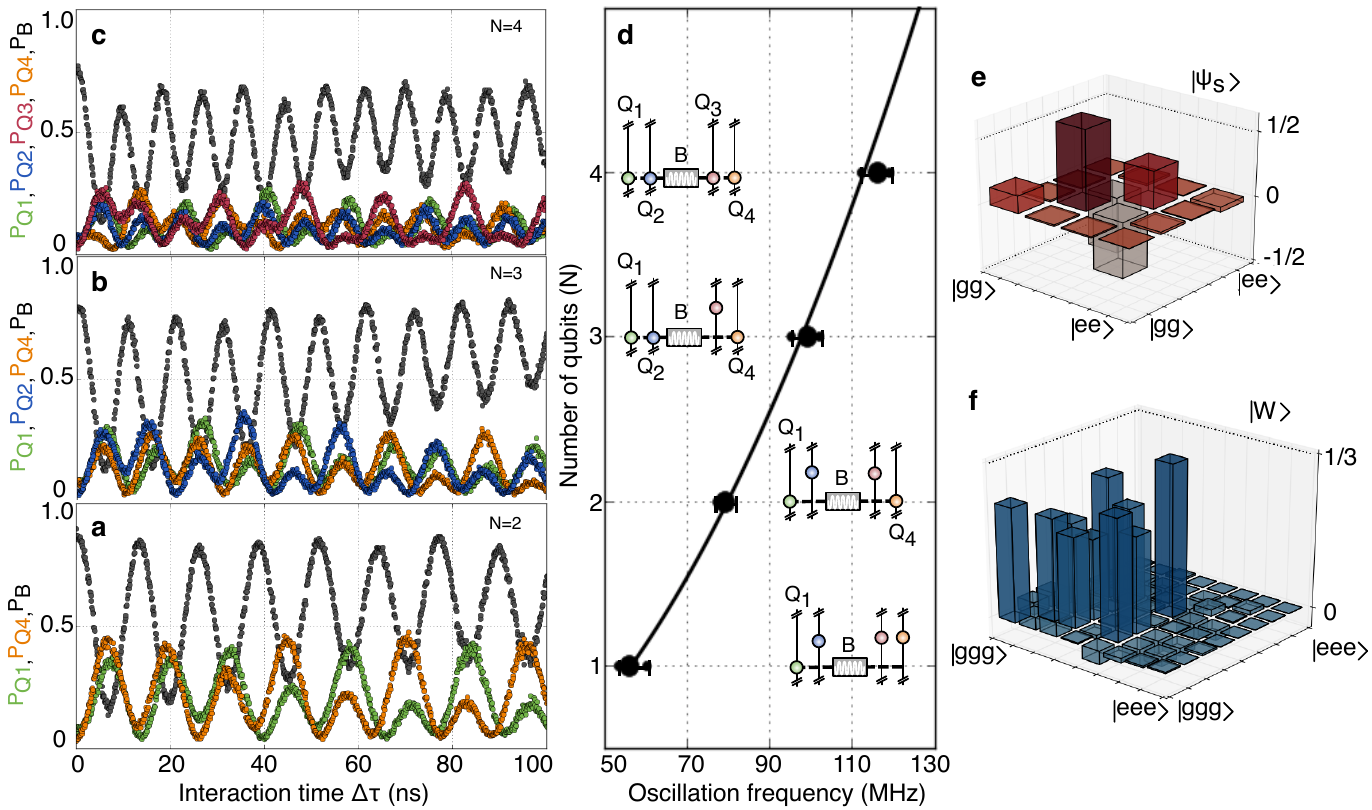}
\caption{\label{timedomain} Rapid entanglement for two to four-qubits. Panels {\bf a,b,c} show the measured state occupation probabilities $P_{Q_{1-4}}$(color) and $P_{B}$(black) for increasing number of participating qubits $N=\{2,3,4\}$ versus interaction time $\Delta\tau$. In all cases B is first prepared in the $n=1$ Fock state\cite{Hofheinz2008} and the participating qubits are then tuned on resonance with B for the interaction time $\Delta\tau$. The single excitation begins in B, spreads to the participating qubits, and then returns to B. These coherent oscillations continue for a time $\Delta\tau$ and increase in frequency with each additional qubit. {\bf d}, Oscillation frequency of $P_{B}$ for increasing numbers of participating qubits. The error bars indicate the $-3\,\mathrm{dB}$ point of the Fourier transformed $P_{B}$ data. The inset schematics illustrate which qubits participate. The coupling strength increases as $\bar{g}_N=\sqrt{N}\bar{g}$, plotted as a black line fit to the data, with $\bar{g}=56.5\pm0.05\,\mathrm{MHz}$. {\bf e,f} The real part of the reconstructed density matrices from QST. ({\bf e}), Bell singlet $\ket{\psi_{s}}=(\ket{ge}-\ket{eg})/\sqrt{2}$ with fidelity $F_{Bell}=\bra{\psi_{s}}\rho_{Bell}\ket{\psi_s}=0.89\pm0.01$ and $\mathrm{EOF}=0.70$. ({\bf f}), Three-qubit $\W=(\ket{gge}+\ket{geg}+\ket{egg})/\sqrt{3}$ with fidelity $F_{W}=\bra{W}\rho_{W}\W=0.69\pm0.01$. The measured imaginary parts (not shown) are found to be small, with ({\bf e}) $|\Im\,\rho_{\mathrm{\psi_{s}}}| < 0.05$ and ({\bf f}) $|\Im\,\rho_{\mathrm{W}}| < 0.06$, as expected theoretically.}
\ifapsmode
\end{figure*}
\else
\end{figure}
\fi
The qubit-resonator interaction can be described by the Jaynes-Cummings model Hamiltonian\cite{Jaynes1963} $H_{\rm{int}}=\sum_{i}(\hbar g_{i}/2)(a^{\dagger}\sigma_{i}^{-}+a\sigma_{i}^{+})$, where $g_{i}$ is the coupling strength between the bus resonator B and the qubit $\mathrm{Q_{i}}$, $a^{\dagger}$ and $a$ are respectively the photon creation and annihilation operators for the resonator, $\sigma_{i}^{+}$ and $\sigma_{i}^{-}$ are respectively the qubit $\mathrm{Q_{i}}$ raising and lowering operators, and $\hbar = h/2\pi$. The dynamics during the interaction between the $i=\{1,2,3,4\}$ qubits and the bus resonator are shown in Fig.\ref{sample}c, and Fig.\ref{timedomain}\,a,\,b,\,c respectively.

For these interactions the qubits $\mathrm{Q_{1}}-\mathrm{Q_{4}}$ are initialized in the ground state $\ket{gggg}$ and tuned off-resonance from B at an idle frequency $f\sim6.6\,\mathrm{GHz}$. $\mathrm{Q_{1}}$ is prepared in the excited state $\ket{e}$ via a $\pi$-pulse. B is then pumped into the first Fock state $n=1$ by tuning $\mathrm{Q_{1}}$ on resonance ($f\sim6.1\,\mathrm{GHz}$) for a duration $1/2g_{1}=\tau\sim9\,\mathrm{ns}$, long enough to complete an iSWAP operation between $Q_1$ and B, $\ket{0}\otimes\ket{eggg}\rightarrow\ket{1}\otimes\ket{gggg}$ \cite{Hofheinz2008}. 
 
The participating qubits are then tuned on resonance ($f\sim6.1\,\mathrm{GHz}$) and left to interact with B for an interaction time $\Delta\tau$. Figures\,\ref{timedomain}\,a,b,c show the probability $P_{Q_{i}}$ of measuring the participating qubits in the excited state, and the probability $P_{B}$ of B being in the $n=1$ Fock state, versus $\Delta\tau$. At the beginning of the interaction the excitation is initially concentrated in B ($P_{B}$ maximum) then spreads evenly between the participating qubits ($P_{B}$ minimum) and finally returns back to B, continuing as a coherent oscillation during this interaction time.

When the qubits are simultaneously tuned on resonance with B they interact with an effective coupling strength $\bar{g}_N$ that scales with the number $\mathrm{N}$ of qubits as $\sqrt{N}$\cite{Fink2009}, analogous to a single qubit coupled to a resonator in a n-photon Fock state\cite{Hofheinz2008}. For N qubits, $\bar{g}_N = \sqrt{N}\bar{g}$, where $\bar{g}=[1/N(\sum_{i=1,N}g^{2}_{i})]^{1/2}$. The oscillation frequency of $P_{B}$ for each of the four cases $i=\{1,2,3,4\}$ is shown in Fig.\ref{timedomain}\,d. These results are similar to Ref.\cite{Fink2009}, but with a larger number N of qubits interacting with the resonator, we can confirm the $\sqrt{N}$ scaling of the coupling strength with N. From these data we find a mean value of $\bar{g}=56.5\pm0.05\,\mathrm{MHz}$.

By tuning the qubits on resonance for a specific interaction time $\tau$, corresponding to the first minimum of $P_{B}$ in Fig.\,\ref{timedomain}a,\,b, we can generate Bell singlets $\ket{\psi_{S}}=(\ket{ge}-\ket{eg})/\sqrt{2}$ and $\W$ states $\W=(\ket{gge}+\ket{geg}+\ket{egg})/\sqrt{3}$. Stopping the interaction at this time ($\tau_{Bell}=6.5\,\mathrm{ns}$ and $\tau_{W}=5.1\,\mathrm{ns}$) leaves the single excitation evenly distributed among the participating qubits and places the qubits in the desired equal superposition state similar to the protocol in Ref.\cite{Altomare2010}. We are able to further analyze these states using full QST. Figures\,\ref{timedomain}e,\,f show the reconstructed density matrices from this analysis\cite{Steffen2006a}. The Bell singlet is formed with fidelity $F_{Bell}=\bra{\psi_{s}}\rho_{Bell}\ket{\psi_s}=0.89\pm0.01$ and entanglement of formation\cite{Hill1997} $\mathrm{EOF}=0.70$. The three-qubit $\W$ state is formed with fidelity $F_{W}=\bra{W}\rho_{W}\W=0.69\pm0.01$, which satisfies the entanglement witness inequality $F_{W}>2/3$ for three-qubit entanglement \cite{Acin2001}.  Generating either of these classes of entangled states (bi- and tri-partite) requires only a single entangling operation that is short relative to the characteristic time for two-qubit gates ($t_{g}\sim50\,\mathrm{ns})$. This entanglement protocol has the further advantage that it can be scaled to an arbitrary number of qubits.



\ifapsmode
\begin{figure*}
\else
\begin{figure}
\fi
\includegraphics[scale=0.95, trim=0in 0in 0in 0in, clip=false]{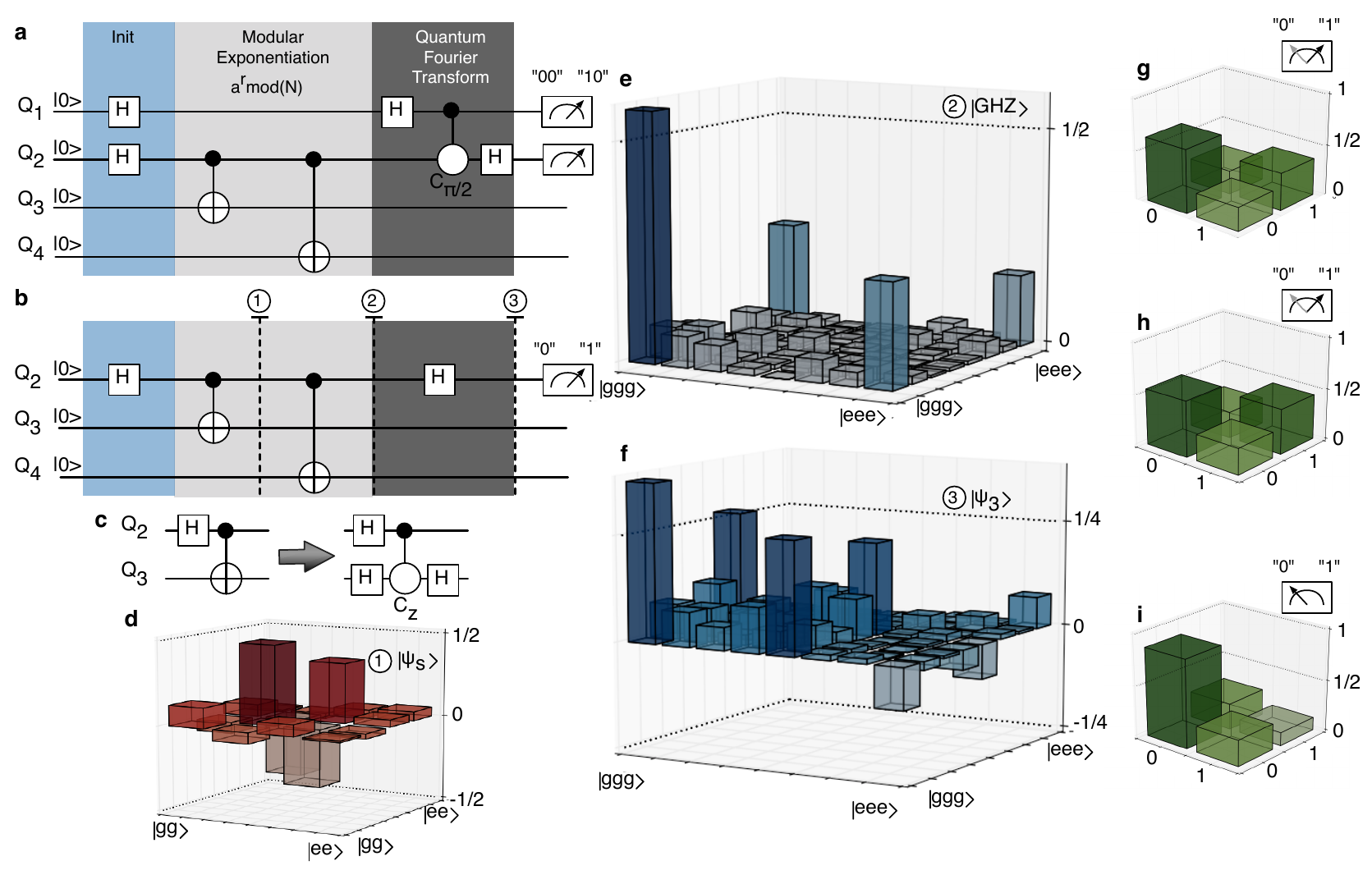}
\caption{\label{shordata} Compiled version of Shor's algorithm. {\bf a}, Four-qubit circuit to factor $\mathrm{N}=15$, with co-prime $a=4$. The three steps in the algorithm are initialization, modular exponentiation, and the quantum Fourier transform, which computes $a^{r}\rm{mod}(N)$ and returns the period $r=2$. {\bf b}, ``Recompiled"  three-qubit version of Shor's algorithm. The redundant qubit $\mathrm{Q_1}$ is removed by noting that $\mathrm{H}\mathrm{H}=\mathrm{I}$. Circuits {\bf a} and {\bf b} are equivalent for this specific case. The three steps of the runtime analysis are labeled 1,2,3. {\bf c}, CNOT gates are realized using an equivalent controlled-Z ($\mathrm{C_Z}$) circuit. {\bf d}, Step 1: Bell singlet between $\mathrm{Q_2}$ and $\mathrm{Q_3}$ with fidelity, $F_{Bell}=\bra{\psi_{s}}\rho_{Bell}\ket{\psi_s}=0.75\pm0.01$ and $\mathrm{EOF}=0.43$. {\bf e}, Step 2:  Three-qubit $\GHZ=(\ket{ggg}+\ket{eee})/\sqrt{2}$ between $\mathrm{Q_2}$, $\mathrm{Q_3}$, and $\mathrm{Q_4}$ with fidelity $F_{GHZ}=\bra{GHZ}\rho_{GHZ}\GHZ=0.59\pm0.01$. {\bf f}, Step 3: QST after running the complete algorithm. The three-qubit $\GHZ$ is rotated into $\ket{\psi_{3}}=H_{2}\GHZ=(\ket{ggg}+\ket{egg}+\ket{gee}-\ket{eee})/2$ with fidelity, $F=0.55$. {\bf g},{\bf h} The density matrix of the single-qubit output register $\mathrm{Q_2}$ formed by: ({\bf g}), tracing-out $\mathrm{Q_3}$ and $\mathrm{Q_4}$ from {\bf f}, and ({\bf h}) directly measuring $\mathrm{Q_2}$ with QST, both with $F=\sqrt{\rho}\,\sigma_{m}\,\sqrt{\rho}=0.92\pm0.01$ and $S_{L}=0.78$. From $1.5\times10^{5}$ direct measurements the output register returns the period $r=2$, with probability $0.483\pm0.003$, yielding the prime factors $3$ and $5$. ({\bf i}), The density matrix of the single-qubit output register without entangling gates, $H_{2}H_{2}\ket{g}=I\ket{g}$. The algorithm fails and returns $\mathrm{r}=0$ $100\,\%$ of the time. Compared to the single quantum state $\ket{\psi_{out}}=\ket{g}$, the fidelity $F_{check}=\bra{\psi_{g}}\rho_{check}\ket{\psi_{g}}=0.83\pm0.01$, which is less than unity due to the energy relaxation. }
\ifapsmode
\end{figure*}
\else
\end{figure}
\fi

The quantum circuit for the compiled version of Shor's algorithm is shown in Fig.\,\ref{shordata}a for factoring the number $N=15$ with  $a=4$ co-prime \cite{Beckman1996, Buscemi2011}, which returns the  period $r=2$ (``$10$" in binary) with a theoretical success rate of $50\,\%$. Although the success of the algorithm hinges on quantum entanglement, the final output is ideally a completely mixed state, $\sigma_{m}=(1/2)(\ket{0}\negthinspace\bra{0}\negthinspace+\negthinspace\ket{1}\negthinspace\bra{1})$. Therefore, measuring only the raw probabilities of the output register does not reveal the underlying quantum entanglement necessary for the success of the computation.  Thus, we perform a runtime analysis with QST at the three points identified in Fig.\ref{shordata}b, in addition to recording the raw probabilities of the output register.

The first breakpoint in the algorithm verifies the existence of bipartite entanglement. A Bell-singlet $\ket{\psi_{s}}$ is formed after a Hadamard-gate ($H$) \cite{Lucero2010} on $\mathrm{Q_{2}}$ and a Controlled-NOT (CNOT)\cite{Yamamoto2010,Mariantoni2011} between $\mathrm{Q_{2}}$ and $\mathrm{Q_{3}}$. Figure\,\ref{shordata}d, is the real part of the density matrix reconstructed from QST on $\ket{\psi_{s}}$. The singlet is formed with fidelity $F_{Bell}=\bra{\psi_{s}}\rho_{Bell}\ket{\psi_s}=0.75\pm0.01$ ($|\Im\,\rho_{\mathrm{\psi_{s}}}| < 0.05$ not shown) and entanglement of formation $\mathrm{EOF}=0.43$\cite{t1}. 

The algorithm is paused after the second CNOT gate between $\mathrm{Q_{2}}$ and $\mathrm{Q_{4}}$ to check for tripartite entanglement. At this point a three-qubit $\GHZ=(\ket{ggg}+\ket{eee})/\sqrt{2}$, with fidelity $F_{GHZ}=\bra{GHZ}\rho_{GHZ}\GHZ=0.59\pm0.01$ ($|\Im\,\rho_{\mathrm{GHZ}}| < 0.06$ not shown) is formed between $\mathrm{Q_{2}}$, $\mathrm{Q_{3}}$, and $\mathrm{Q_{4}}$ as shown in Fig.\ref{shordata}e. This state is found to satisfy the entanglement witness inequality, $F_{GHZ}>1/2$ \cite{Acin2001} indicating three-qubit entanglement.

The third step in the runtime analysis captures all three qubits at the end of the algorithm, where the final $H$-gate on $\mathrm{Q_{2}}$, rotates the three-qubit $\GHZ$ into $\ket{\psi_{3}}=H_{2}\GHZ=(\ket{ggg}+\ket{egg}+\ket{gee}-\ket{eee})/2$. Figure\,\ref{shordata}f is the real part of the density matrix with fidelity $F=\bra{\psi_{3}}\rho_{3}\ket{\psi_{3}}=0.54\pm0.01$. From the three-qubit QST we can trace out the register qubit to compare with the experiment where we measure only the single qubit register and the raw probabilities of the algorithm output. 
Ideally, the algorithm returns the binary output``$00$" or ``$10$" (including the redundant qubit) with equal probability, where the former represents a failure and the latter indicates the successful determination of $r=2$. We use three methods to analyze the output of the algorithm: Three-qubit QST, single-qubit QST, and the raw probabilities of the output register state. Figures\,\ref{shordata}g,\,h are the real part of the density matrices for the single qubit output register from three-qubit QST and one-qubit QST with fidelity $F=\sqrt{\rho}\,\sigma_{m}\,\sqrt{\rho}=0.92\pm0.01$ for both density matrices. From the raw probabilities calculated from ${150{,}000}$ repetitions of the algorithm, we measure the output ``$10$" with probability $0.483\pm0.003$, yielding $r=2$, and after classical processing we compute the prime factors $3$ and $5$.

The linear entropy $S_L=4[1-\mathrm{Tr}(\rho^{2})]/3$ is another metric for comparing the observed output to the ideal mixed state, where $S_L=1$ for a completely mixed state\cite{White2007}. We find $S_{L}=0.78$ for both the reduced density matrix from the third step of the runtime analysis (three-qubit QST), and from direct single-qubit QST of the register qubit. 

As a final check of the requisite entanglement, we run the full algorithm without any of the entangling operations and use QST to measure the single-qubit output register. The circuit reduces to two $H$-gates separated by the time of the two entangling gates. Ideally $\mathrm{Q_{2}}$ returns to the ground state and the algorithm fails (returns ``$0$") $100\,\%$ of the time. Figure\,\ref{shordata}i is the real part of the density matrix for the register qubit after running this check experiment. The fidelity of measuring the register qubit in $\ket{g}$ is $F_{check}=\bra{g}\rho_{check}\ket{g}=0.83\pm0.01$. The algorithm fails, as expected, without the entangling operations. 

In conclusion, we have implemented a compiled version of Shor's algorithm on a QuP that correctly finds the prime factors of 15. We showed that the QuP can create Bell states, both classes of three-qubit entanglement, and the requisite entanglement for properly executing Shor's algorithm. In addition, we produce coherent interactions between four qubits and the bus resonator with a protocol that can be scaled to create an $N$-qubit $\W$ state, during which we observe the $\sqrt{N}$ dependence of the effective coupling strength with the number N of participating qubits. These demonstrations represent an important milestone for superconducting qubits, further proving this architecture for quantum computation and quantum simulations.

Devices were made at the UCSB Nanofabrication Facility, a part of the NSF-funded National Nanotechnology Infrastructure Network. This work was supported by IARPA under ARO awards W911NF-08-1-0336 and W911NF-09-1-0375. R.B. acknowledges support from the Rubicon program of the Netherlands Organisation for Scientific Research. M.M. acknowledges support from an Elings Postdoctoral Fellowship. The authors thank S. Ashhab and A. Galiautdinov for useful comments on rapid entanglement.

Correspondence and requests for materials should be addressed to J.M.M.~(email: martinis@physics.ucsb.edu).

\bibliography{shor}

\begin{thebibliography}{10}
\bibitem{t1}
 \bibinfo{note}{We find energy relaxation and dephasing from the qubits as the
  primary source of reduced fidelities. $T_{1}\sim400\mathrm{ns}$ and
  $T_{2}\sim200\mathrm{ns}$ for all four qubits and $T_{1}\sim3\mathrm{\mu s}$
  for the bus resonator.}
\end{thebibliography}

\end{document}